# Multilayer silicene: structure, electronics, and mechanical property

Chen Qian,[1] Zhi Li*[2]

1. Department of mechanical engineering, Zhejiang University, 866 Yuhangtang Rd, Hangzhou 310058, P.R. China.

2. College of Material Science and Engineering, Nanjing University of Science and Technology, Nanjing 210094, China

Herein, we performed first principle calculation and classical molecular dynamics simulation to study structural optimization, band structure, and mechanical properties of differently stacked multilayer silicene. Several local energy minima have been identified as metastable conformation with different stacking mode and layer number. Bandstructure of low buckled AA bilayer silicene optimized with SCAN+rvv10 presents semiconducting behavior with a bandgap of 0.4419ev. Young's modulus of multilayer silicene shows low dependency on layer number or stacking mode. Whereas, fracture stress and strain is sensitive to the number of layers, specific stacking mode, and chirality. Furthermore, bending modulus of multilayer silicene (e.g., 0.44ev for monolayer silicene) is even lower than that of graphene, which may attribute to the flexibility of bond angle.

## 1. Introduction

With the rise of 2D material, the honeycomb structure of group iv element has attracted tremendous attention due to its exotic electronic, optical, thermal, and mechanical property. Silicene, the counterpart of graphene with buckled honeycomb structure have spurred increasing interest since it was epitaxially grown on Ag(111) due to its compatibility with silicon-based electronics[1]. For the silicene structure deposited on Ag(111)[1], Ag(110)[2], Ir(111)[3], zirconium diboride thin film[4], atoms in the honeycomb structure displace vertically, leading to different silicene superstructures at different temperature[5]. Scalable free-standing silicene preparation is achieved by liquid oxidation and peeling of CaSi2[6] recently and found with potential application in lithium-ion batteries.

Silicene presents a different structure from graphene, which exhibits out-of-plan bucking length about 0.40Å while graphene is planar. The buckling distance is a result of $sp^2$-$sp^3$ hybridization of silicon atoms. The bandstructure of pristine monolayer silicene shows a zero bandgap with a Dirac cone at K point in Brillouin zone[7] and has been confirmed by experiment[8]. Since the zero bandgap of silicene impedes its application in semiconductor industry, several strategies have been proposed to open the bandgap. The bandgap of monolayer silicene opens and increases with the magnitude of vertical electrical filed by changing the buckling distance[9, 10], which make silicene a potential candidate for field-effect transistors[11]. Several possible stacking orders of free-standing bilayer silicene has been studied, while AB' stacking bilayer silicene is a direct semiconductor with approximately 0.29ev bandgap[12] and slide-2AA is an indirect bandgap semiconductor with 1.16ev which is quite close to the bandgap of bulk silicon(1.1ev)[13]. The bandgaps of oxidized $\sqrt{13} \times \sqrt{13}$, $4 \times 4$, and $2\sqrt{3} \times 2\sqrt{3}$ superstructure grown on Ag(111) are 0.18, 0.9, 0.22ev respectively considering higher chemical reactivity of silicene than graphene[14]. Doping by different types of transition metal elements with varying coverage is also confirmed to be a feasible routine of tuning the bandgap of silicene[15].

While graphene present extremely high mechanical strength and Young's modulus, the mechanical property of silicene remain one of the interests since the first time it was reported. Fracture dynamics of silicene membrane were investigated by DFTB and ReaxFF, plane stiffness were 43.0 N.m$^{-1}$(for both ACM and ZZM with ReaxFF), 62.7 N.m$^{-1}$ for ACM and 63.4 N.m$^{-1}$ for ZZM with DFTB, the buckling magnitude is alleviated during loading process[16]. Moreover, free-standing monolayer was found to be mechanical unstable at room temperature due to the dangling bond at the edge[17] and will be stabilized through hydrogenation at the side of the monolayer silicene nanosheets[7]; meanwhile, bilayer silicene sheet remains stable at ambient temperature[17]. Mechanical property of planar and buckled bilayer silicene structure through molecular dynamics simulation found a transition from buckled structure to planner structure during tensile process[18].

Since most works to date focus on monolayer silicene and properties and structures of multilayer silicenes remains to be explored, in this article structure optimizations and band structures are investigated through ab initio method while bending modulus and tensile process are studied by massive parallel MD simulations for silicenes from 1 to 4 layers. In contrast to multilayers graphene in which carbon atoms from different layers interact with van der Waals force, silicon atoms of multilayer silicene are covalently bonded to adjacent layers. The AA bilayer silicene is predicted to be an indirect bandgap semiconductor with 0.4419ev bandgap. Bending modulus of multilayer silicene is quite different from the case for graphene owing to the buckled structure with flexible bond angle and the strong covalent bond between layers which suppresses the interlayer shear during bending.

## 2. Computational details

To construct multilayer silicene, we introduce three fundamental stacking orders to depict the relative position between two adjacent layers, which are AA, AA' and AB stacking order[12, 13]. As illustrated in Fig. 1, AA stacking order is composed of two same layers of silicon atoms with a translation along the vertical direction. Atoms in the second layer of AA' stacking displace toward the reverse direction of the first layer with covalent bond connecting the upward atoms in the first layer and the downward atoms in the second layer. AB stacking order is an analog of multilayer graphene but with buckled structure leading to the covalent bond between fractional atoms of two layers like AA' stacking. Multilayer structures are named after these stacking orders in this article.

In the study of structure optimization for multilayer silicene of different stacking orders, strongly constrained and appropriately normed (SCAN) functional[19] and long-range vdW interaction from rVV10 considered (SCAN+rVV10) functional[20] are employed to compare the cohesive energy of different lattice morphologies. SCAN is the first meta-GGA that satisfies all known constraints and capable of capturing much of intermediate-range vdW interaction through exchange term with almost the same degree of the computational cost of GGA functional. A 15Å thick vacuum space in the direction perpendicular to atomic plan is used to eliminate the effect of the periodic image. During the process of structural optimization, K points in hexagonal Brillouin zone are sampled with $25 \times 25 \times 1$ Monkhorst-Pack mesh. Cutoff energy for the plane wave basis set is 400ev. Gaussian smearing method is adopted with sigma equal to 0.01, while $1 \times 10^{-5}$ev is specified for electronic self-consistent loop and ionic relaxation loop will stop after all the force on the atom is smaller than $1 \times 10^{-4}$ev/Å .

The HSE06 hybrid functional[21] is employed to explore the band structure for multilayer silicene within local minimum during structure optimization. In HSE06, a screened Coulomb potential term is used for HF exchange interaction to accelerate computation and range-separation parameter equal to 0.2 is selected. The exchange-correlation energy is separated into 25% of short-range Hartree-Fock exchange energy, 75% of short-range PBE short-range exchange energy, long-range PBE exchange energy, and PBE correlation energy. To acquire the whole electronic structure in reciprocal space, a Brillouin zone constituted with $45 \times 45 \times 1$ Monkhorst-Pack mesh is used. Brillouin zone consists of $15 \times 15 \times 1$ Monkhorst-Pack mesh and 18, 11 and 21 k-points with 0 weight selected along $\Gamma \to M \to K \to \Gamma$ respectively is constructed to study the band structure along high symmetry route. Both structure optimization and band structure calculation are performed by Vienna Ab Initio Simulation Package(VASP)[22] and projected augment wave (PAW) method[23]. The density of charge is visualized with VESTA[24].

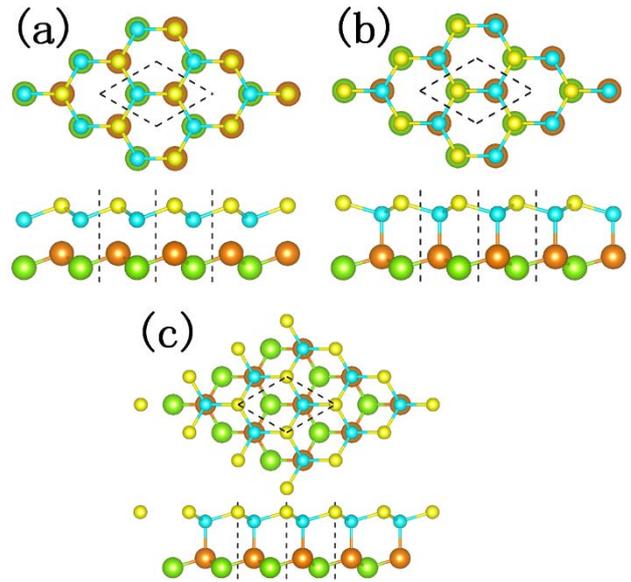

**Figure. 1** Top and side view of three $3 \times 3$ basic stacking mode: (a) AA; (b) AA'; (c) AB. Atoms with larger radius correspond to bottom layer. Different colours denote atoms in different buckling directions.

In molecular dynamics simulation of Bending modulus and tensile simulation, reaxFF[25] is used to describe the interaction between silicon atoms. The reaxFF is derived from the training set of first principle calculation and capable of modeling the bond breaking and forming process with bond length and bond order considered[26]. During the calculation of bending modulus, $10\text{nm} \times 10\text{nm}$ silicene is firstly used for energy minimization with periodic boundary along atomic plan under zero external pressure. Silicene after first energy minimization runs is rolled into cylinders with different curvatures. Atoms within the margin of 5Å alongside each edge of the circumference are fixed. Energy minimization is then performed again for the silicene cylinder and the energy of unfixed atoms is used for fitting of bending modulus. All the energy minimization runs are implemented by conjugate gradient method with convergence criterion of $10^{-8}$ ev/Angstrom.

For the tensile simulation, $10\text{nm} \times 10\text{nm}$ silicene under periodic boundary condition for planar direction is used to simulate the dynamic tensile process of infinite silicene. Silicene is thermalized for 20ps with time step of 0.2fs under 10K temperature and 0 pressure after energy minimization to exclude initial stress before tensile process. Control of temperature and pressure (NPT) is achieved using Nose-Hoover thermostat and barostat with temperature and stress damping parameter of 20 and 200 fs to eliminate excessive thermal and pressure fluctuation. Strain rate during the uniaxial tensile process is $10^{-9}$/s for the desired tensile direction while pressure of the other direction is kept around 0. Considering the ambiguity of size of silicon atoms, thickness of each silicene layer is considered to be the van der Waals diameter of Si, i.e.

4.2Å. Stress versus strain data from the first 5% train stage is used for calculation of Young's modulus. Simulations of both chiralities (armchair and zigzag) are considered.

## 3. Results and discussion

**Part I: geometry and electronic structure of multilayer silicene**

The optimized low-buckled monolayer silicene present lattice constant of 3.84Å, bond length of 2.25Å, and buckling distance of 0.40Å, which is in good agreement with the previous study[7]. With the three basic stacking orders introduced, multilayer silicene structures are constructed as combinations of these tacking orders. The diversity of the stacking orders is a result of the buckled structure of single layer silicene. All the multilayer silicene conformations studied in this article are named with these three basic tacking orders, e.g. AA'B' trilayer silicene means that the first layer and the second layer are stacked with AA' stacking mode while the second layer and the third layer are stacked with AB stacking mode.

One dimensional cohesive energy is scanned as a function of lattice constant at the range of 2.5Å to 4.5Å for each structure. Cohesive energy of different structure with changing lattice constant is illustrated in Fig. 2. Local energy minima of each conformation correspond to locally stable structures. Both SCAN and SCAN+rvv10 show the same energy tendency and almost same structure during optimizing, indicating that long-range vdW interaction is negligible for multilayer silicene. A number of phase transitions are observed for a few initial structures with the increasing lattice constant and will be discussed later.

For single layer silicene, two local minimum is observed in Fig. 2a, corresponding to high buckled structure and low buckled structure reported in the previous studies[7]. High buckled silicene has higher energy ($\Delta_E$=0.15348ev), longer bond length ($\Delta_l$=0.35Å) and larger buckling distance ($\Delta_\Delta$=1.68Å) compared with low buckled structure. Atoms of high bucked structure tend to cluster for larger supercell[7], indicating instability of which. Therefore, high buckled structure only exists under the constraint of single unit cell and is less considered in existing literature. The low buckled structure is related to covalent bond with sp$^2$ and sp$^3$ hybridization as a compromise for lower energy of sp$^3$ hybridization than sp$^2$ one under the constraint of the honeycomb structure, leading to highly chemical active feature, unlike graphene. The bond angle of low bucked one is 116.86 degree, which falls in between that of sp$^2$ hybridization ($\theta = 120°$) and sp$^3$ hybridization ($\theta = 109°28'$). Doping of Li$^+$ can suppress the pseudo-Jahn-Teller distortion (PJT) and give rise to a planar structure with 1.62ev band gap[27]. Based on the consideration of the instability of monolayer high buckled structure, we mainly focus on the low buckled (LB) multilayer silicene conformation with lattice constant between 3.5Å and 4.5Å.

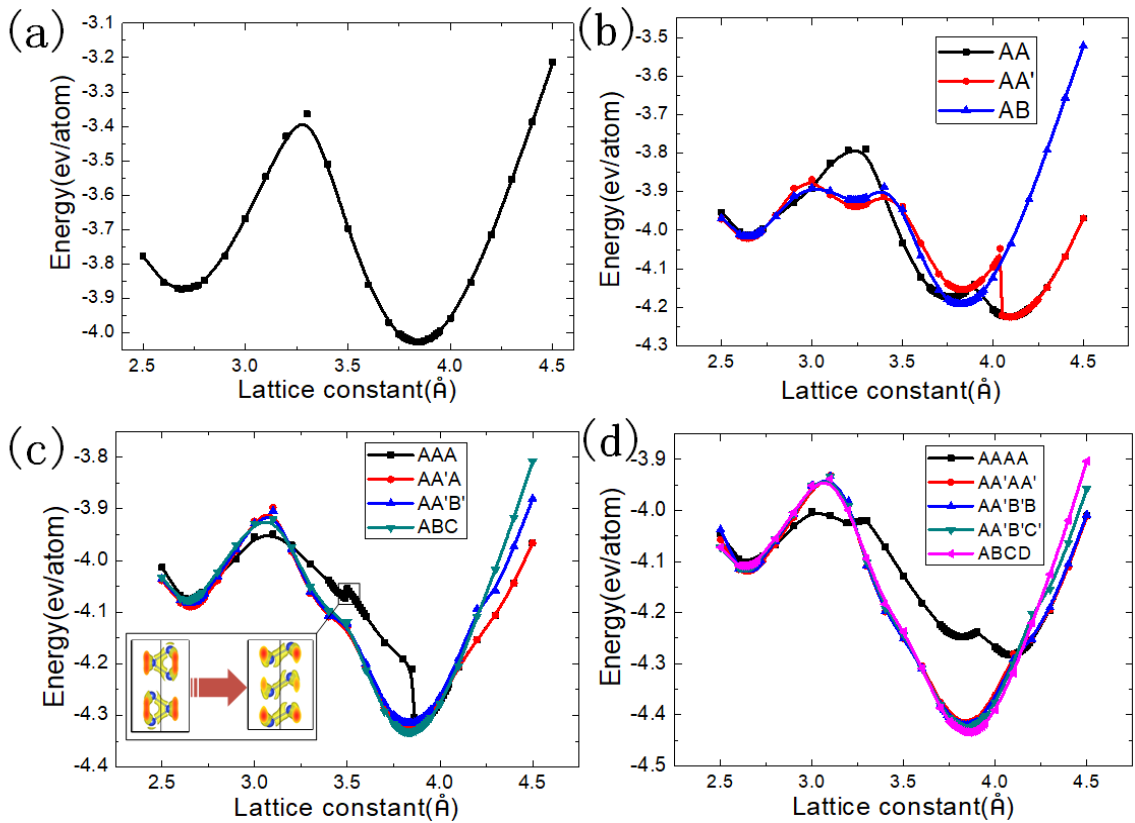

**Figure. 2** Negative value of cohesive energy per Si atoms as a function of lattice constant for: (a) 1 layer; (2) 2 layer; (3) 3 layer; (4) 4layer. All the values are calculated with SCAN+rvv10. Inset in (c) illustrate charge density of bilayer 2H molybdenum disulphide like structure transiting to three typical AA stacked layers.

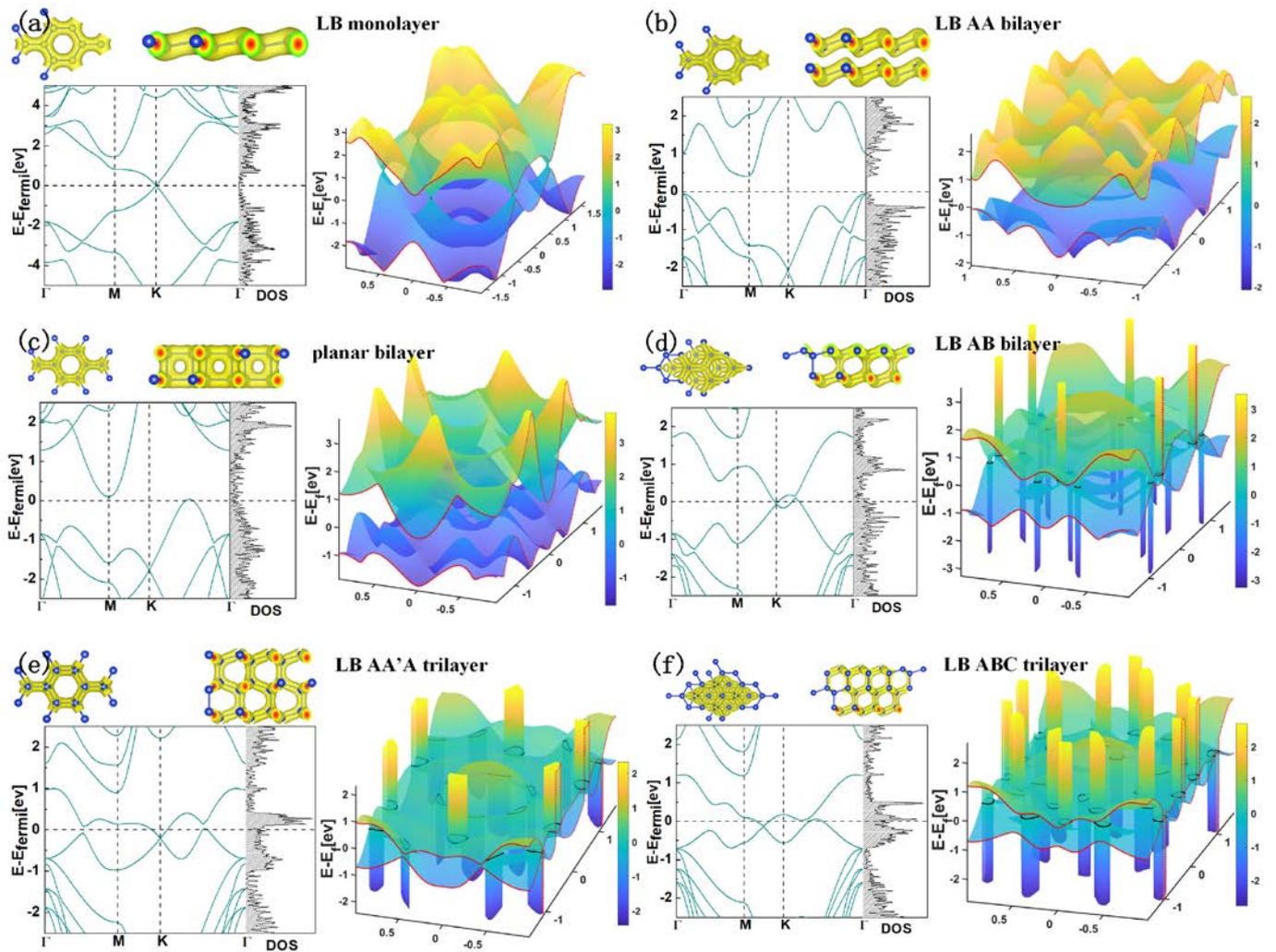

**Figure. 3** top and side view of Charge density (upper left panel); band structure along Γ→M→K→Γ of Brillouin zone and overall density of states (DOS) (lower left panel); 3D plot of LUMO band and HOMO band as a function of K-points in Brillouin zone (right panel) for: (a) monolayer; (b) LB AA bilayer; (c) planar bilayer; (d) LB AB bilayer; (e) LB AA'A trilayer; (f) LB ABC trilayer.

In the case of bilayer silicene, three structures (AA, AA' and AB) are studied in Fig. 2b. Two local minimum is noticed between 3.5Å and 4.5Å for AA type, one with low bucked structure at 3.76Å and the other with planar bilayer structure at 4.09Å. Atoms from different sublayers of AA low buckled structure interact with van de Waals force instead of covalent bond. AA planar structure consists of two planar honeycomb sublayers like graphene. All the atoms from different sublayers of AA planar structure are covalently bonded in the vertical direction. During the optimization of AA' structure, AA' stacked low buckled structure is energy minimized at 3.83Å and abruptly transit toward planar structure after 4.05Å and finally reach a local minimum at 4.09Å with the same bilayer planar structure as in AA case. Besides, a transition from AA' low buckled bilayer structure to planar bilayer structure has been observed during tensile simulation under canonical (NVT) ensemble[28]. For AB, one locally stable structure at 3.82Å is noticed and has the lowest energy over these three stacking modes around the neighborhood of 3.8Å. All the bilayer stacking modes possess local minimum around 2.7Å and 3.8Å, correspond to the high buckled and low buckled structure of monolayer silicene while many other local minima are also observed. The planar bilayer structure presents the lowest energy among all the bilayer conformations discussed, indicating that it's the most stable bilayer structure.

In Fig. 2c, energies of trilayer AA'A, AA'B' and ABC structure present similar tendencies with the evolutional lattice constant, while AAA trilayer shows an entirely different one. For all the AA', AA'B' and ABC structure, two local minimum is located on the 1D energy surface, one located around 2.7Å (2.65Å for AA', 2.65Å for AA'B', 2.64Å for ABC) with high buckled sublayers and the other around 3.8Å (3.84Å for AA', 3.83Å for AA'B', 3.83 for ABC) with low buckled sublayers and much lower energy. However, AA trilayer presents two energy discontinuity at 3.5Å and 3.87Å. At the lattice constant of 3.49Å, the upward buckled atom of the second sublayer interact with three nearest neighbor atom site in the third sublayer through covalent bond while the downward buckled atom in the second sublayer is covalently bonded to three nearest neighbor atom site in the first sublayer, resembling the conformation of bilayer 2H molybdenum disulfide. As the lattice constant increases to 3.50Å, atoms from different sublayers interact with each other with vdW force instead. The transition from 3.49Å AAA trilayer to 3.50Å AAA trilayer is illustrated in the inset of Fig. 2c. Typical AAA trilayer transits to AA'A trilayer structure after 3.87Å for lower energy.

The tendency of energy in respect to the lattice constant for tetralayer silicene is similar to that of trilayer structure. Structures except AAAA tetralayer, show two local minima like trilayer ones. The phase transition from typical AA stacked four layer silicene to a planar tetralayer structure is observed. Typical AAAA stacked tetralayer reaches to local minimum energy at 3.82Å. Atoms from different low bucked sublayers of AAAA tetralayer interact with vdW force. The planar tetralayer consist of two planar bilayers, i.e., the first sublayer atoms are covalently bonded to the second sublayer, the second sublayer interact with the third sublayer through vdW force, the third sublayer and the fourth sublayer are covalently bonded like the first sublayer and second layer.

Among all the structure probed, low buckled structure (around 3.8Å) and high buckled structure (around 2.7Å) are observed, corresponding to the high bucked and low bucked structures of monolayer silicene. The buckled distance of low bucked structure increases with the number of layers, indicating larger sp$^3$ component for multilayer silicene. Cohesive energy of multilayer silicene increases with the number of layers as a result of decreased surface energy and finally approaches to that of bulk silicon (4.63Å). The energy difference between low buckled structure and high buckled structure also increases with the number of layers, which verifies the stability of low bucked multilayer. Structural parameters and cohesive energy of the LB multilayer are listed in supplementary material.

In consideration of layer-dependent bandgap of MoS2[29] and phosphorene[30], electronic structures of different layers and stacking modes are probed. In Fig. 3, the upper left panel denotes the charge density of the corresponding structure, from which we distinguish covalent bond between atoms. The 2D band structures diagram along $\Gamma \to M \to K \to \Gamma$ direction is presented on the lower panel of each graphene. Besides the 2D band structure, the surface of the energy of lowest unoccupied molecular orbital (LUMO) and highest occupied molecular orbital (HOMO) at each K-points is plotted at the right panel. Different from 2D band structure plot, we should note that the surface of energy of LUMO and HOMO may discontinue at particular region when two bands which are the closest two bands from Fermi energy in the area crossing each other under or over the Fermi energy, making parts of other bands the LUMO or HOMO energy within that region. The black lines in Fig. 3 stand for the intersection of LUMO or HOMO with Fermi energy, thicker black line means smaller slope of the original band crossing Fermi energy. The 3D band structure is necessary for the study of the entire electronic property of 2D materials and useful to exclude Dirac cone like structure in 2D band structure plot like the case of LB AA' bilayer[12]. In Fig. 3a, we can tell the Dirac cone located at the K point of primitive hexagonal Brillouin zone with π and π* band crossing fermi level linearly for monolayer LB silicene. The massless Dirac fermions like charge carrier and ambipolar character can be predicted for single layer silicene from the 2D and 3D band structure plot, which has been reported in previous work[7]. In Fig. 3c, the LUMO and HOMO band of bilayer planar morphology encounter fermi level tangentially but don't intersect with each other, making it a semi-metal. For LB AA'A trilayer, a Dirac cone locates at K point below Fermi energy like n-type doped. Dirac cone below Fermi level also exist in the case of LB ABC trilayer, but is located in somewhere between M and K points as illustrated in Fig. 3f. In Fig. 3d, two very shallow Dirac cones exist in LB AB bilayer, one below Fermi energy at K point and the other above Fermi energy at somewhere between K and Γ points, but both of the Dirac cones are too shallow to notice. The band gap of 0.4419ev opens in LB AA bilayer, making it an indirect bandgap semiconductor. Compared with LB AA bilayer studied in previous study[12, 13] with metallic electronic structure while the structure of which is optimized with LDA or GGA functional, the meta-GGA (SCAN and SCAN+rvv10) optimized structure is more trustworthy. Unfortunately, no further semi-conductor is discovered among investigated conformations, as most of the morphologies studied are metallic. Band structures of all the LB morphologies can be found in supplementary material.

**Part II: bending modulus and tensile test of multilayer silicene**

Minimized energy of unfixed atoms for different LB structures bent with curvature at the range of 0.005~0.03Å$^{-1}$ is presented in Fig. 5. The out of plan bending modulus is calculated with the binomial fitting of energy versus curvature to the following formula:

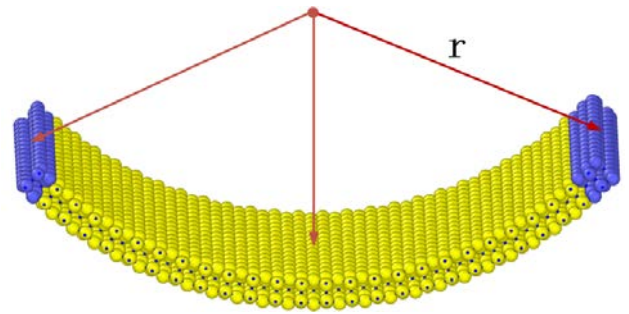

**Figure. 4** Illustration of bending simulation for silicene with imposed radius of curvature.

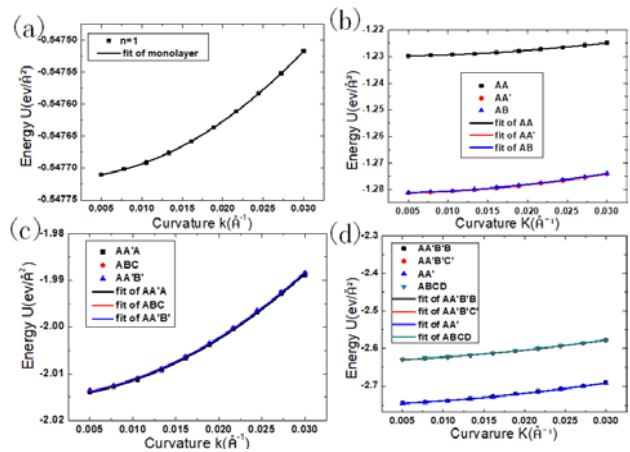

**Figure. 5** Energy as a function of curvature for different stacking order with number of layers (a) n=1; (b) n=2; (c) n=3; (d) n=4.

$$U = \frac{1}{2}Dk^2$$

where U is strain energy per unit area of the basal plane (ev/Å$^2$), D is bending modulus per unit width (ev), and k is the prescribed beam curvature of the bent silicene sheet (Å$^{-1}$) in Fig. 4. Bending modulus are listed in Table. 1 for different morphologies. Compared with graphene[31] (bending moduli of 1 Layer 2 Layer 4 Layer types are 2.1ev 130ev and 1199ev respectively), silicene has remarkably smaller bending modulus. The smaller bending modulus of silicene can be explained with higher flexibility of bond angle and relatively higher energy related with sp$^2$-sp$^3$ hybridization of silicene. The mechanism of the flexibility of the bond angle is discussed in supplementary material. AA bilayer has smaller bending modulus (11.2125ev) compared with AA' (16.30044ev) and AB (16.399ev). The difference between AA and AA' or AB bilayer is expectable, since covalent bonds between sublayers of AA' or AB impede interlayer shear effect, leading to larger bending modulus. For four layer, the small difference between AA'B'B or AA' and AA'B'C' or ABCD is merely a result of different lattice sizes in consideration of the fact that the values of energy per atom after energy minimization for the four morphologies are nearly the same. With the participation of interlayer covalent bond which impedes shear between layers during bending process, bending modulus of silicene is expected to be more sensitive to the number of layers than graphene in actual experiment[32].

**Table 1** Bending modulus (ev) of multilayer silicene with varying number of layers and stacking mode.

| Number of layers | Morphologies | Bending modulus (ev) | R$^2$ |
|---|---|---|---|
| N=1 | monolayer | 0.43688 | 0.99999 |
| N=2 | AA | 11.2125 | 1 |
|  | AA' | 16.30044 | 1 |
|  | AB | 16.399 | 1 |
| N=3 | AA'A | 50.61782 | 0.99974 |
|  | ABC | 50.99538 | 0.99975 |
|  | AA'B' | 50.8934 | 0.99974 |
| N=4 | AA'B'B | 81.17372 | 0.9991 |
|  | AA'B'C' | 78.50912 | 0.99911 |
|  | AA'AA' | 80.99572 | 0.9991 |
|  | ABCD | 78.28936 | 0.9991 |

As is well known to us, the mechanical property of material is significantly related to the atomic structure. Attempts are made to compare the mechanical difference between monolayer and multilayer silicene. The stress versus strain curve is plotted in Fig. 6. Young's modulus, fracture stress and fracture strain in armchair and zigzag directions are determined from the strain-stress relationship and listed in Table. 2. For all these structures, fracture stress in zigzag direction is larger than that in armchair direction. The discrepancy mainly attributed to the larger bond angle deviation during the tensile process in zigzag direction[16] while bond angle distortion accounts for part of strain besides bond stretching. In Fig. 6b, fracture stress of AB is higher than AA' in both directions. Moreover, significantly higher strain before the final rupture of AA' bilayer in zigzag direction is noticed. With strain approaching the value of 0.244 in zigzag direction, bilayer AA' transits to planer structure, leading to higher fracture stain. Whereas in armchair direction, the phase transition to planar bilayer is not observed, possibly related to the less bond angle distortion. For trilayer in Fig. 6(c), ABC trilayer presents the highest strength in zigzag direction and AA' in armchair direction. Strength of AA'B' falls between that of AA' and ABC. Note that the sharp change of stress for AA'A trilayer in armchair direction at strain around 0.2 corresponds to a sudden shrink of lateral magnitude. Furthermore, the abrupt stress change of AA'A trilayer in zigzag direction around strain of 0.224 is in accordance with the sudden expansion in armchair direction. As shown in Fig. 6c, ABCD tetralyer has the largest strength in zigzag direction and AA'AA' tetralayer in armchair direction. Similar sharp stress change for several morphologies in Fig. 6c is observed with analogical mechanism for trilayer. Above all, monolayer silicene presents higher fracture stress and strain. Fracture stress and strain of multilayer silicene seems to be more dependent on specific stacking mode and chirality as well as number of layers.

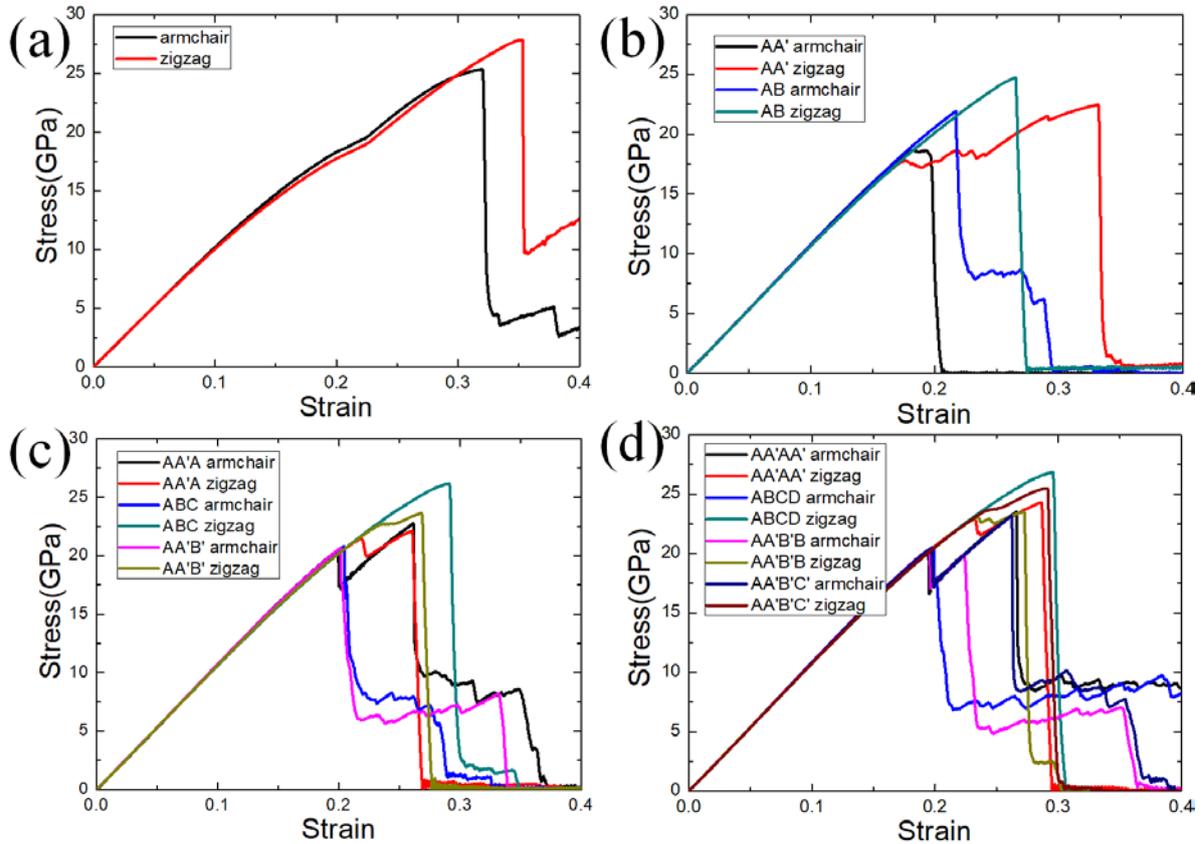

**Figure. 6** Comparison of stress versus strain curve for each chirality of differently stacked (a) monolayer; (b) bilayer; (c) trilayer; (d) tetralayer.

Next, we look into the effect of number of layers on Young's modulus. Young's modulus is obtained from stress versus strain data by Hooke's law $\sigma = E\varepsilon$. Only tiny increment (merely about 3.5%) is noticed from single layer to two layers and the marginal difference between 2-4 layer silicene is negligible. Moreover, Young's modulus in armchair and zigzag direction for each morphology is approximately equal. Young's modulus is also independent of stacking mode. As listed in Tabl.e 2, Young's modulus of investigated silicene locate in the range of 103.2Gpa is higher than AA' in both directions. Moreover, significantly and 107.63Gpa, which is merely 0.1 of graphene (1.09Tpa ~1.13Tpa when number of layer is between 1 and 7) [33].

## 4. Conclusions

To summarize, multilayer silicene constructed from AA, AA' ad AB stacking mode is used to study structure optimization, band structure and mechanical property by first principle calculation and classical molecular dynamics simulation. Energy difference between optimized lowest energy and bulk silicon eliminates with increase of layer number. The bucking length is larger for thicker silicene, indicating larger component of sp$^3$ hybridization. Charge density map of AAA trilayer with a small lattice constant of 3.49Å shows a bilayer 2H molybdenum disulfide type conformation. LB AA bilayer is predicted to present semi-conducting behavior with a bandgap of 0.4419ev while other multilayer silicene conformations are metallic. Young's modulus of multilayer silicene with different stacking mode and layer number shows marginal difference like graphene[33]. However, fracture strength and fracture strain show strong dependency on morphologies including stacking mode and layer number of multilayer silicene. Zigzag direction generally exhibits higher strength than armchair direction for multilayer silicene with the participation of bond angle distortion. Moreover, bending modulus for multilayer silicene is even lower than graphene, possibly resulting from the high flexibility of the bond angle.

## Conflicts of interest

There are no conflicts to declare.

## Acknowledgements

We are grateful to National Supercomputer Center in Guangzhou for high performance computational resource.

**Table 2** Comparison of Young's modulus (E), fracture stress ($\sigma_f$), fracture strain ($\epsilon_f$) in each chirality of different layer number and stacking order

|   |   | chirality | E (GPa) | $\sigma_f$ (GPa) | $\epsilon_f$ |
|---|---|---|---|---|---|
| N=1 | \ | armchair | 103.05 | 25.45 | 0.32 |
|   |   | zigzag | 103.02 | 27.90 | 0.35 |
| N=2 | AA' | armchair | 106.65 | 18.65 | 0.18 |
|   |   | zigzag | 106.55 | 22.46 | 0.33 |
|   | AB | armchair | 107.14 | 21.94 | 0.22 |
|   |   | zigzag | 106.56 | 24.68 | 0.27 |
| N=3 | AA'A | armchair | 107.11 | 22.93 | 0.26 |
|   |   | zigzag | 107.60 | 22.26 | 0.26 |
|   | ABC | armchair | 106.60 | 20.99 | 0.20 |
|   |   | zigzag | 107.37 | 26.37 | 0.29 |
|   | AA'B' | armchair | 107.56 | 20.84 | 0.20 |
|   |   | zigzag | 107.14 | 23.83 | 0.27 |
| N=4 | AA'AA' | armchair | 107.43 | 23.51 | 0.27 |
|   |   | zigzag | 107.52 | 24.32 | 0.29 |
|   | ABCD | armchair | 107.00 | 20.55 | 0.20 |
|   |   | zigzag | 106.98 | 26.86 | 0.30 |
|   | AA'B'B | armchair | 107.52 | 19.93 | 0.22 |
|   |   | zigzag | 107.63 | 23.48 | 0.27 |
|   | AA'B'C' | armchair | 107.02 | 23.19 | 0.26 |
|   |   | zigzag | 106.86 | 25.5 | 0.29 |

# Supplementary information

**Table S1** Optimized structural parameters and cohesive energy of multilayer slicene with low buckled and planar morphology. Symbols of b, h, Δ and Ec represent lattice constant, distance between layers, buckling length and cohesive energy calculated with SCAN+rvv10 respectively.

|  | b(Å) | h(Å) | Δ(Å) | $E_c$(ev) |
|---|---|---|---|---|
| LB monolayer | 3.84 | / | 0.40 | 4.03 |
| LB AA bilayer | 3.76 | 2.78 | 0.82 | 4.17 |
| Planar AA bilayer | 4.09 | 2.39 | 0 | 4.23 |
| LB AA' bilayer | 3.83 | 3.10 | 0.64 | 4.15 |
| LB AB bilayer | 3.82 | 3.16 | 0.64 | 4.19 |
| LB AA' 3 layer | 3.84 | 3.12 | 0.66/0.75 | 4.33 |
| LB AA'B' 3 layer | 3.83 | 3.15/3.10 | 0.67/0.76/0.65 | 4.31 |
| LB AB 3 layer | 3.83 | 3.13 | 0.67/0.75 | 4.33 |
| LB AA 3 layer | 3.82 | 2.81/3.03 | 0.65/0.77 | 4.25 |
| Planar AA 4 layer | 4.08 | 2.42/3.09 | 0 | 4.28 |
| LB AA' 4 layer | 3.84 | 3.12/3.14 | 0.64/0.77 | 4.42 |
| LB AA'B'B 4layer | 3.84 | 3.12/3.12 | 0.65/0.77 | 4.42 |
| LB AA'B'C' 4layer | 3.85 | 3.12/3.11/3.06 | 0.65/0.77/0.76/0.62 | 4.42 |
| LB AB 4layer | 3.85 | 3.07/3.11 | 0.62/0.76 | 4.43 |

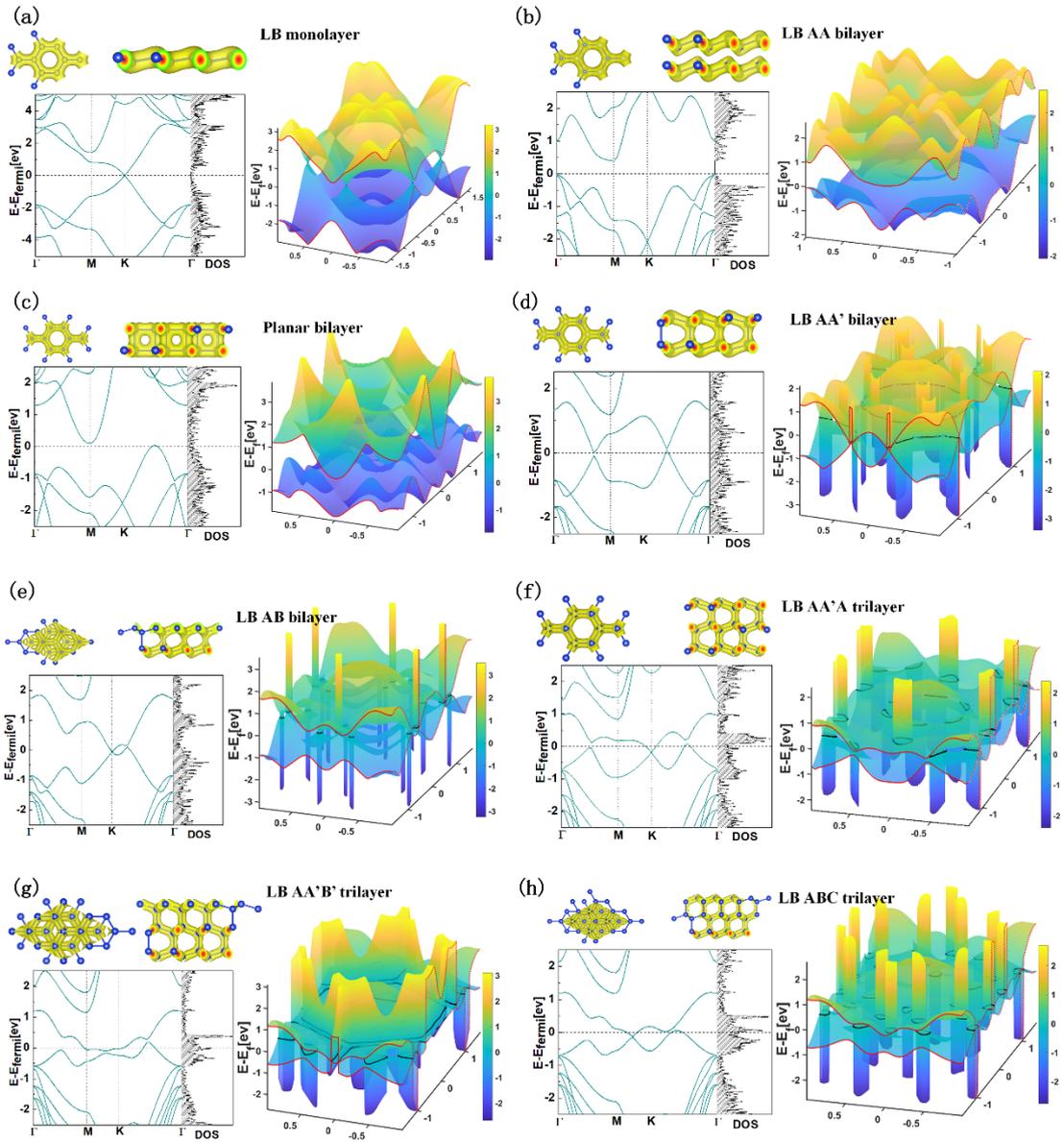

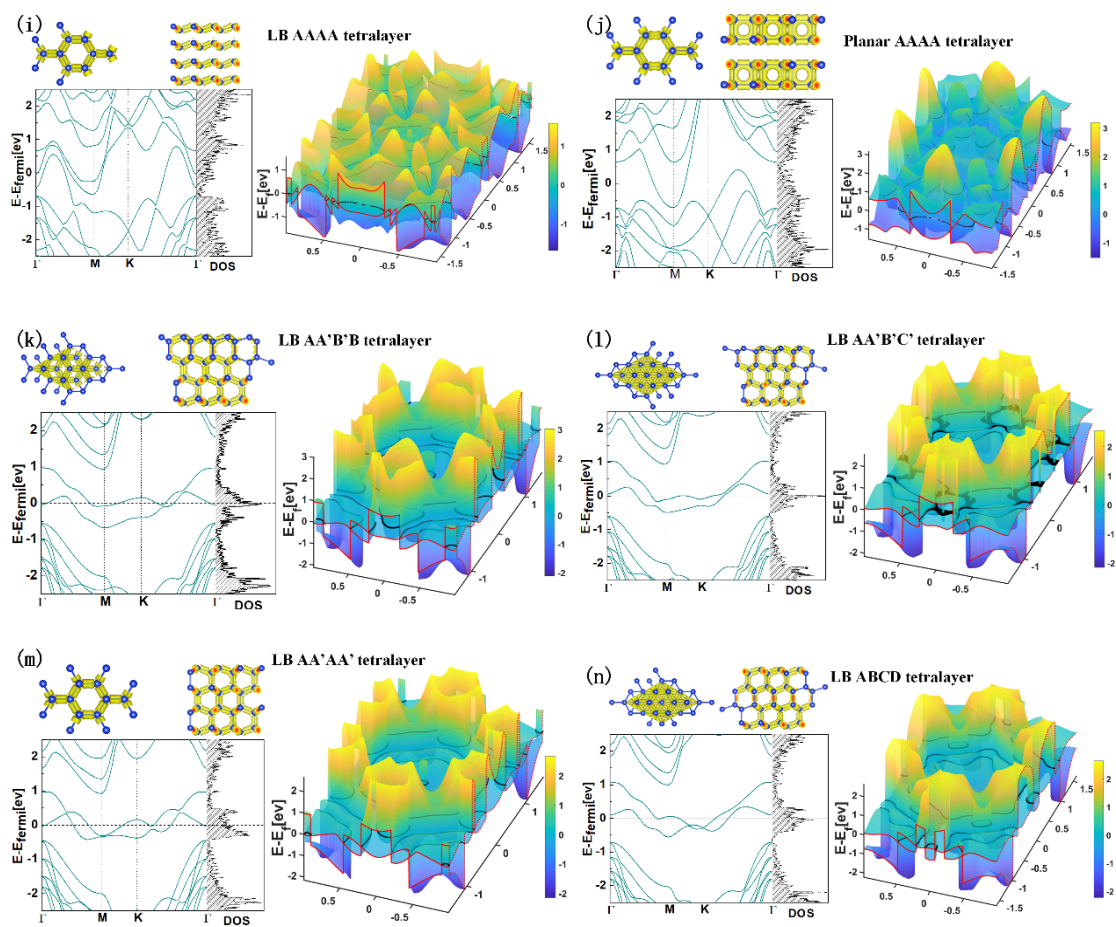

**Figure S1.** Charge density of $3 \times 3$ supercell (upper left panel), band structure along $\Gamma \to M \to K \to \Gamma$ route of Brillouin zone and overall density of states (DOS) (lower left panel), 3D plot of LUMO band and HOMO band as a function of K-points in Brillouin zone (right panel).

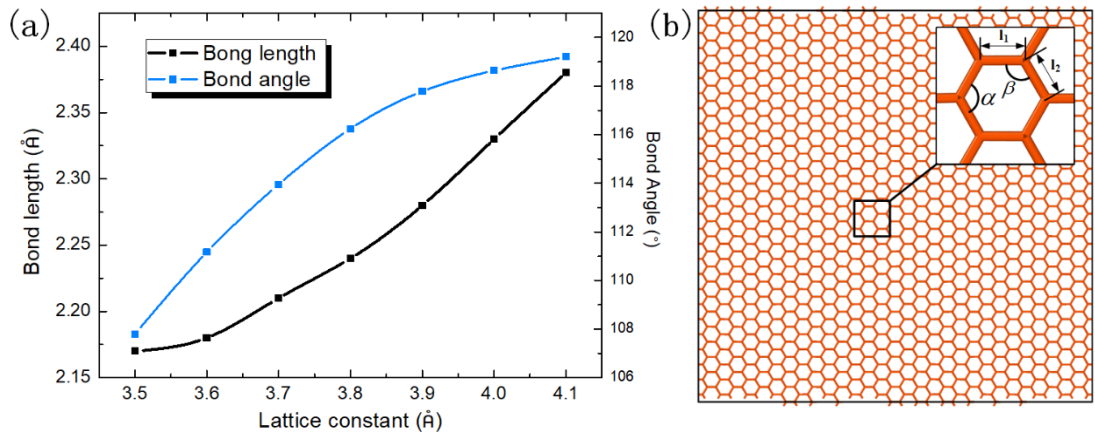

**Fig S2.** Illustration of flexibility of bond angle: (a) bond length and bond angle of monolayer silicene with varying lattice constant during structure optimization (b) α, β bond angle and $l_1$, $l_2$ bond with different circumstance during tensile process

The flexibility of bond angle and bond length deviation with varying lattice constant is illustrated in Fig S2. As Lattice constant increases from 3.5Å to 4.1Å, not only bond length but also bond angle contributes to the deviation of lattice constant. There are two different bond lengths ($l_1$, $l_2$) and bond angles (α,β) during tensile process in different loading direction. With denser alignment of α in zigzag direction, α will suffer from higher deviation and suppress the elongation of $l_2$ during loading in zigzag direction, resulting in higher strength in zigzag direction. More detailed analysis about evolvement of $l_1$, $l_2$, α, β during tensile process of monolayer silicene can be found in reference article[1].